\documentclass{JHEP3}
\def\be{\begin{equation}}
\def\ee{\end{equation}}
\def\bea{\begin{eqnarray}}
\def\eea{\end{eqnarray}}
\def\const{\textrm{const}}

\preprint{DAMTP-2005-56 \\ hep-th/0506222}
\title{Three Charge Supertubes in Type IIB Plane Wave Backgrounds}
\author{Hari K. Kunduri and James Lucietti \\  DAMTP, Centre of Mathematical
Sciences, Cambridge University,\\ Wilberforce Road, Cambridge CB3
0WA, UK \\ \email{H.K.Kunduri@damtp.cam.ac.uk} \qquad
\email{J.Lucietti@damtp.cam.ac.uk }} \abstract{We deform the
supersymmetric black ring of five
  dimensional supergravity coupled to $N-1$ vector multiplets
  to obtain an asymptotically G\"odel supersymmetric black ring. For the $U(1)^3$ model we
  lift this solution to obtain a three charge D1-D5-P supertube
  which asymptotes to a 1/2 supersymmetric plane wave of Type
  IIB supergravity. Further, we also show how one may deform the asymptotically flat three charge supertube of type IIB, in the special case of vanishing KK dipole charge, to a three charge supertube which asymptotes to the maximally supersymmetric plane wave.}
\keywords{Black rings, G\"odel spacetime, supertubes, Plane waves}
\begin{document}

\section{Introduction}
Following the discovery of the vacuum rotating
black ring\footnote{Although another rotating black ring has recently
  been found~\cite{Pau}, due to its conical singularity it is strictly not a
  vacuum solution.}\cite{EmpRea} a number of generalisations have been
  constructed. These could be organised broadly into two groups. The
  first encompasses asymptotically flat charged rotating rings, which
  include both the supersymmetric~\cite{minsusyring} and non-extremal black rings~\cite{nonextrem} of
  minimal supergravity as well as the dipole rings found in~\cite{Dip}. The
  second set consists of non-asymptotically flat solutions. It is
  relatively straightforward to use solution-generating techniques to construct
  black rings in fluxbranes~\cite{tubular,magnetic, elec}.  More interesting has
  been the recent construction of supersymmetric black rings in
  Taub-NUT backgrounds~\cite{elv,bena,stro}. Furthermore, Ortin has made use of the
  fact one can deform supersymmetric solutions of minimal supergravity
  in order to derive a black ring that asymptotes to the maximally
  supersymmetric G\"odel spacetime~\cite{Ortin}. In this note we show how one can
  trivially generalise this procedure to the more general case of
  minimal supergravity coupled to $N-1$ abelian vector multiplets. While
  these solutions are interesting in their own right, we will be
  interested in using them as a means to construct supertube
  configurations in ten and eleven dimensions.
\par
One can lift solutions of the $U(1)^3$ five dimensional supergravity
  to eleven dimensions. When one does this with the supersymmetric
  black ring solution, one obtains a black supertube. Via dimensional
  reduction and a series of certain T-dualities, they represent
  D1-D5-P supertubes in Type IIB supergravity~\cite{3chargestube,onering,GauGut}. On the other hand,
 performing this sequence of dualities on the G\"odel solution, one
  obtains a supersymmetric plane wave~\cite{BGHV}. Thus one would expect~\cite{Ortin} that lifting G\"odel black rings would lead to supertubes
  embedded in a plane wave. Indeed a three charge G\"odel BMPV black
  hole was constructed in~\cite{brecher} using the reverse procedure. In
  that work, an asymptotically plane wave D1-D5-P configuration was
  constructed, which upon T-duality and dimensional reduction led to a
  three-charge G\"odel BMPV black hole. In this note we construct three charge black supertubes that are
  asymptotically plane wave. We find that the near horizon geometry is
  unchanged by the deformation. These generalise the supertubes
  of~\cite{3chargestube}. We find the presence of the Kaluza-Klein
  monopole tube somewhat obstructs the inclusion of the plane
  wave. More precisely, if the Kaluza Klein dipole charge $q^3$ is non zero,
  our solutions necessarily contain closed timelike curves. Further, we
  can only construct a three charge supertube in the maximally supsersymmetric
  plane wave in the case where $q^3 = 0$. 
\par This paper is organised as follows. First, we review the
  construction and supersymmetric solutions of the general five
  dimensional minimal supergravity coupled to $N-1$ abelian vector
  multiplets. We show how these can be deformed simply. Next, we
  uplift the solution and present the resulting Type IIB
  configuration. We show explicitly that far from the branes it
  asymptotes to a supersymmetric plane wave, supported by the appropriate
  Ramond-Ramond three form flux. In the following section we consider
  embedding the three charge supertubes in more general supersymmetric
  plane wave backgrounds. In particular we derive a solution with
  three charges and two dipole charges that
  asymptotes to the maximally supersymmetric plane wave. Finally, we
  conclude with a discussion and comment on extensions of the work.

\section{Asymptotically G\"odel supersymmetric black rings}
To begin we will concern ourselves with $D=5$ minimal supergravity
coupled to $N-1$ abelian vector multiplets with scalars valued in a
symmetric space. We follow the notation of~\cite{3chargestube}. The
action of this theory is: \bea\label{fiveaction} S= \frac{1}{16\pi
}\int \left( R *1 - G_{IJ} dX^I \wedge *dX^J - G_{IJ} F^I \wedge
*F^J -\frac{1}{6} C_{IJK}F^I \wedge F^J \wedge A^K \right) \; \eea
where $I,J,K=1,...,N$ and the scalars $X^I$ are constrained by \be
\frac{1}{6} C_{IJK}X^I X^J X^K  =1 \ee where $C_{IJK} = C_{(IJK)}$
and the following condition is obeyed:
\begin{equation}
C_{IJK}C_{J'(LM}C_{PQ)K'}\delta^{JJ'}\delta^{KK'} =
\frac{4}{3}\delta_{I(L}C_{MPQ)}.
\end{equation}
The matrix $G_{IJ}$ is defined as \be G_{IJ} =\frac{9}{2} X_I X_J
-\frac{1}{2}C_{IJK}X^K \ee where one lowers indices on the scalars
as follows: \be X_I = \frac{1}{6} C_{IJK}X^J X^K. \ee The
classification of supersymmetric solutions of this theory can be
deduced from that of the gauged theory initiated in~\cite{GutRea}
and subsequently generalized in~\cite{GutSab}. Such solutions admit
a globally defined non-spacelike Killing vector field $V$. If there
exists a neighbourhood where $V$ is timelike one can choose
coordinates $(t,x^m)$ such that $V=\partial/\partial t $ and \be
\label{fivemetric} ds^2 = -f^2(dt+\omega)^2 +f^{-1} h \ee where $h$
is a Riemmannian metric on the base-space $\mathcal{B}$, $f$ is a
function and $\omega$ a 1-form both living on $\mathcal{B}$. Let
$e^0 =f(dt+\omega)$ and we will choose an orientation of
$\mathcal{B}$ such that $e^0 \wedge \textrm{vol}(h)$ is positively
oriented. We decompose $d\omega$ into self-dual and anti self-dual
parts on the base as \be fd\omega =G^+ + G^-. \ee Supersymmetry then
implies that $h$ is a hyper-K\"ahler metric on $\mathcal{B}$ and
that \be \label{fivemaxflux} F^I= d(X^Ie^0) +\Theta^I \ee where
$\Theta^I$ are self-dual two forms on $\mathcal{B}$ such that \be
X_I\Theta^I =-\frac{2}{3}G^+. \ee The Bianchi identities for $F^I$
then give $d\Theta^I=0$. The Maxwell equations imply \be \nabla^2
(f^{-1}X_I) = \frac{1}{6}C_{IJK}\Theta^J \cdot \Theta^K, \ee where
$\alpha \cdot \beta \equiv \frac{1}{p!} \alpha^{m_1m_2 \dots \, m_p}
\beta_{m_1m_2 \dots \, m_p}$ for $p$-forms $\alpha$ and $\beta$ on
$\mathcal{B}$.

 Ortin has made the interesting observation that a
solution of the minimal theory can be deformed by adding a piece to
$G^-$ while still leaving it a solution. This obviously generalises
to the case considered here where we have a minimal theory coupled
to $N-1$ vector multiplets. He showed that by a judicious choice of
$G^-$ one can make the supersymmetric ring asymptotically G\"odel.
We show here that the same construction works for the supersymmetric
ring of the $U(1)^N$ theory.

Firstly, we write down the $U(1)^N$ supersymmetric black ring. The
base $(\mathcal{B},h)$ is $\mathbb{E}^4$. We write the metric in toroidal coordinates as
\begin{equation} \label{base}
h = \sum_{i=1}^4 dx^2_i = \frac{R^2}{(x-y)^2}\left ((y^2-1)d\psi^2 +
\frac{dy^2}{y^2 -1} +
 (1-x^2)d\phi^2 + \frac{dx^2}{1-x^2}  \right).
\end{equation} The various quantities are given by
\bea \label{quantities}
\Theta^I &=& -\frac{1}{2}q^I(dy \wedge d\psi +dx \wedge d\phi), \\
f^{-3} &=& \frac{1}{6}C^{IJK}H_IH_JH_K, \\ \nonumber
\frac{1}{3} H_I &\equiv& f^{-1}X_I = \bar{X}_I + \frac{1}{6R^2} \left(
Q_I -\frac{1}{2}C_{IJK}q^Jq^K \right)(x-y) \\ && \qquad \qquad -\frac{1}{24R^2} C_{IJK}
  q^Jq^K (x^2 -y^2), \\ \label{lastring}
\omega &=& \omega_{\phi} d\phi +\omega_{\psi} d\psi, \\
\omega_{\phi} &=& -\frac{1}{8R^2}(1-x^2)[ q^IQ_I -\varrho(3+x+y)],
\nonumber \\ 
\omega_{\psi} &=& \frac{3}{2}(1+y)q^I \bar{X}_I-\frac{1}{8R^2}(y^2-1)[
  q^IQ_I -\varrho(3+x+y)], \nonumber
\eea where $q^I$, $Q_I$, $\bar{X}_I$ are constants,
$C^{IJK}=C_{IJK}$ and $\varrho= \frac{1}{6} C_{IJK} q^Iq^Jq^K$. The
constants $\bar{X}_I$ obey the same constraint as $X_I$ do. The
coordinate ranges are, as usual for black rings, $-1 \leq x \leq 1$
and $-\infty < y \leq -1$ and both angles have period $2\pi$. There
is an event horizon at $y = -\infty$. Using the method of Ortin, we
deform the solution as follows. Simply replace $\omega$ by $\omega'=
\omega + \omega_G$ where $\omega_G$ is given by \be \omega_G =
\frac{\mu R^2}{(x-y)^2} [ (1-x^2) d\phi -(y^2-1) d\psi ]. \ee It is
easy to check that $(d\omega_G)^+=0$\footnote{The orientation is
defined by
  $\epsilon_{y\psi x\phi}=+1$ as in~\cite{3chargestube} and corresponds to $\epsilon_{x_1x_2x_3x_4}=+1$.} and thus
$G^+$ for this deformed solution is the same as in the undeformed
case. This is what implies it is still a solution. The remarkable
fact is that this deformation leaves the horizon intact and thus the
solution still represents a black ring. This is easy to see since as
$y \to -\infty$ the extra terms arising in the metric from
$\omega_G$ vanish. Also, as promised, the solution asymptotes to the
maximally supersymmetric G\"odel spacetime. To see this one needs to
introduce the polar coordinates \be \rho \sin\theta =
\frac{R\sqrt{y^2-1}}{x-y}, \qquad \rho \cos\theta = \frac{R
\sqrt{1-x^2}}{x-y} \ee where $0 \leq \theta \leq \pi /2$ and $0 \leq
\rho < \infty$. Then using the fact that the undeformed solution is
asymptotically flat it is easy to deduce that as $\rho \to \infty$
\be\label{fiveGodel} ds^2 \sim -(dt+\omega_G)^2 + d\rho^2 +\rho^2
(d\theta^2 +\sin^2\theta d\psi^2 +\cos^2\theta d\phi^2). \ee In
these coordinates $\omega_G = \mu \rho^2 \sigma^3_R/2$ and
$\sigma^3_R= d\phi'+\cos\theta'd\psi'$ is a right invariant
form\footnote{The Euler angles $(\theta',\phi',\psi')$ are given by
$\theta'=2\theta$, $\psi'=
  \phi+\psi$ and $\phi'= \phi-\psi$.} on the
$S^3$. This corresponds to the maximally supersymmetric G\"odel
solution. Note that this particular deformation does not introduce
any Dirac-Misner strings ($\omega'_{\phi}(x=\pm 1) =
\omega'_{\psi}(y=-1)=0$) but closed timelike curves (CTC) will of
course occur. We should note that the near-horizon limit of this
deformed black ring is still locally $AdS_3 \times S^2$. This is
easy to see using the technique of~\cite{3chargestube}. The extra
terms present in the metric due to the deformation are \be
-f^2\omega_G^2 -2f^2(dt+\omega)\omega_G \ee and as $y \to
-\infty$, $f^2 \sim \const \times y^{-4}$, $\omega \sim \const
\times y^3 d\psi$ and $\omega_G \sim -\mu R^2 d\psi$. Then
letting\footnote{The radius of the $S^1$ of the ring at the horizon
is denoted by $L$ and is a function of the charges and $R$ as given
in~\cite{3chargestube}.} $\tilde{r}=-R^2/(\epsilon L y)$ and
$\tilde{t}=t/ \epsilon$ it is clear that as $\epsilon \to 0$ both of
the above terms vanish. Note that the extra term present in each of
the gauge fields also vanishes in this limit.

\par The solution we have constructed can therefore be seen to describe asymptotically G\"odel supersymmetric black rings coupled to $N-1$
vector multiplets. As in the undeformed case, they are $\frac{1}{2}$
BPS.  We note in passing that in the $R=0$ limit one recovers the
analogous supersymmetric G\"odel BMPV black hole. This is made
manifest by passing into the $(\rho,\theta)$ coordinate system.

\section{Asymptotically plane wave D1-D5-P supertube}
It is well known that $D=11$ supergravity, reduced on a $T^6$ with
coordinates $z_{a}$, $a=\{1,...,6 \}$, using the Ansatz
\begin{eqnarray}
ds^2_{11} & = & ds_{5}^2 + X^{1}(dz_{1}^2 + dz_{2}^2) +
X^{2}(dz_{3}^2 +
dz_{4}^2) + X^3(dz_{5}^2 + dz_{6}^2) \nonumber \\
C_{3} & = & A^{1} \wedge dz_{1} \wedge dz_{2} + A^{2}\wedge dz_{3}
\wedge dz_{4} + A^{3}\wedge dz_{5} \wedge dz_{6}
\end{eqnarray} yields the STU model. This has the action~(\ref{fiveaction}) with $N=3$,
the only non vanishing component of $C_{IJK}$ is $C_{123}=1$ and
permutations, and the matrix $G_{IJ} = \frac{1}{2}(X^{I})^{-2}
\delta_{IJ}$. In this particular case, the solution  $ds^{2}_{5}$ is
given by~(\ref{fivemetric}) along
with~(\ref{base})-(\ref{lastring}) and the replacement $\omega
\rightarrow \omega'$, describes a $\frac{1}{2}$-BPS three-charge
G\"odel black ring. Note that the field
strengths~(\ref{fivemaxflux}) are also deformed. Lifting these black
rings to eleven dimensional supergravity then yields a
straightforward deformation of the three-charge M-theory supertubes
given in~\cite{3chargestube}. Explicitly, in terms of the $H_{I}$,
we rewrite~(\ref{fivemetric}) as
\begin{equation}
ds_{5}^2 = -(H_{1}H_{2}H_{3})^{-\frac{2}{3}}(dt + \omega')^{2} +
  (H_{1}H_{2}H_{3})^{\frac{1}{3}} \sum_{i=1}^4 dx_i^2
\end{equation} and the 1-form potentials are
\begin{equation}
A^{I} = H_I^{-1}(dt+\omega') - \frac{q^{I}}{2}((1+y)d\psi +
(1+x)d\phi).
\end{equation} Here, as in~\cite{3chargestube} we have set
$\bar{X}_{I}=\frac{1}{3}$. The four supercharges of the five
dimensional solution are inherited to yield $\frac{1}{8}$ BPS
configuration.  The undeformed system presented
in~\cite{3chargestube} consists of three M2 branes carrying
conserved charges proportional to the $Q^{I}$, and three M5 branes,
each of which wrap the `ring' $\psi$ coordinate which is transverse
to the membranes. As explained clearly in~\cite{3chargestube}, these
M5 branes do not carry conserved charges but instead possess
`dipole' charges parameterised by the $q^{I}$. It should be noted
that the notion of mass for these objects is defined relative to an
asymptotic Minkowskian region transverse to the M2 branes. However,
rather than being asymptotically flat, the solution presented here
obviously asymptotes to the $\frac{5}{8}$ BPS supersymmetric G\"odel
universe that arises upon lifting~(\ref{fiveGodel}).

To construct the Type IIB solution, Kaluza-Klein reduce
on the compact direction $z_{6}$ and T-dualize along
$z_{5},z_{4},z_{3}$. We could of course choose any of the $z$ as the
initial $S^{1}$. This choice corresponds to taking $z_{5}$ to point
along the axis of the supertube. The resulting string frame metric
is
\begin{equation}
\label{IIB} ds^2_{\textrm{IIB}} = -\frac{(dt +
\omega')^2}{H_{3}\sqrt{H_{1}H_{2}}} +
 \frac{H_{3}}{\sqrt{H_{1}H_{2}}}(dz_{5}+A^{3})^2 + \sqrt{H_{1}H_{2}} \sum_{i=1}^4 dx_i^2 +
 \sqrt{\frac{H_{2}}{H_{1}}}\sum_{i=1}^{4}(dz_{i})^2
\end{equation} with dilaton
\begin{equation}
e^{2\Phi} = \frac{H_{2}}{H_{1}}
\end{equation} and three-form RR field strength
\begin{equation}
F_{3} = (X^{1})^{-2} \star_{5}F^{1} + F^{2}\wedge(dz_{5} + A^{3}).
\end{equation} The solution above is similar to the
D1-D5-P `double-helix' supertube, except now it is not
asymptotically flat. Given the fact that G\"odel spacetimes are
T-dual to plane waves, we expect something similar to manifest
itself in this solution. To see this explicitly let us study the
asymptotic form of this IIB solution. As $\rho \to \infty$ \be
ds^2_{\textrm{IIB}} \sim dz_5^2 + 2dz_5 (dt+\omega_G) + \sum_{i=1}^4
dx_i^2 + \sum_{i=1}^4 dz_i^2 \ee and if we let $Z=t+z_5$ we get \be
ds^2_{\textrm{IIB}} \sim  -dt^2+dZ^2 + 2(dZ-dt) \omega_G +
\sum_{i=1}^4 dx_i^2 + \sum_{i=1}^4 dz_i^2. \ee Note that this form
of the metric is just as that found in~\cite{BGHV} after T-dualising
the G\"odel IIA solution. Thus if one makes the
coordinate transformation \bea \label{cotr} &&u=t-Z, \qquad v=t+Z \\ \label{coord}
&&\widetilde{\phi}=\phi - \mu u, \qquad \widetilde{\psi} = \psi +\mu
u, \\ &&\widetilde{x}_1+i\widetilde{x}_2 =r_1 e^{i\widetilde{\phi}},
\qquad \widetilde{x}_3+i\widetilde{x}_4 =r_2 e^{i\widetilde{\psi}}
\eea and noting $\rho^2 = r_{1}^2 + r_{2}^2$, the asymptotic form of
the metric becomes \be ds^2_{\textrm{IIB}} \sim -dudv -\mu^2 \left(
\sum_{i=1}^4 \widetilde{x}_i^2 \right) du^2 + \sum_{i=1}^4
d\widetilde{x}_i^2 + \sum_{i=1}^4 dz_i^2 \ee which is a 1/2
supersymmetric plane wave solution to type IIB. The flux becomes \be
F_{3} \sim -\frac{\mu}{2}du \wedge d(\rho^2
\widetilde{\sigma^3_R}).\ee In fact it is the Penrose limit of
$AdS_3 \times S^3 \times T^4$ supported by a three form RR-flux,
which can be derived as the S-dual of $AdS_3 \times S^3 \times T^4$
supported by an NS-NS three form~\cite{russo}. Hence the solution
(\ref{IIB}) seems to represent a D1-D5-P supertube in this plane
wave background. One needs to check that this IIB solution has a
regular horizon. Since we have noted that $ds_5^2$ is regular as $y
\to -\infty$ we simply need to show that $dz_5+A^3$ is also regular.
One can do this in exactly the same manner as was done
in~\cite{3chargestube} by performing the same shift in $z_5$ as they
did. This is because the extra term we have in $A^3$ is
$H_3^{-1}\omega_G$ which vanishes as $y \to -\infty$. In fact one
can go further and show that the near-horizon limit of this IIB
solution is unchanged by the deformation and thus is locally $AdS_3
\times S^3 \times T^4$. This can be deduced from the fact that
$ds_5^2$ has the same near-horizon limit as the undeformed case
together with the fact that the extra term in $A^3$ is
$O(\bar{r}^2)$, where $\bar{r}=-R/y$. Thus the IIB solution we have constructed
interpolates between $AdS_3 \times S^3 \times T^4$ and its Penrose
limit.

A subtlety concerning the global structure of the spacetime derived
here should be noted. Since the Kaluza-Klein direction $z_{5}$ is
compact with period $2\pi R_z$ as in the undeformed solution, the
lightcones coordinates $u=-z_5$ and $v=2t+z_5$ inherit this
periodicity. Indeed, enforcing the coordinate transformation
(\ref{coord}) to be valid globally one deduces that $2 \pi R_z \mu =
2 \pi N$ for some integer $N$. Thus we see that the strength of the
flux of the plane wave $\mu = \frac{N}{R_z}$ is quantized in units
of inverse radius of the compact direction $z_5$. Asymptotically the
deformed solution is actually a discrete quotient of a plane wave.
Hence we will have CTC since $\frac{\partial}{\partial u}$ is
timelike in this plane wave. It is not surprising that we have this
situation, given that before the T-dualities we were dealing with an
asymptotically G\"odel spacetime which contains CTC at every point
for sufficiently large radius. However in the IIB solution the CTC
are milder in the sense that they are global, just like the ones
encountered in AdS spacetimes. Moreover, these features are not
specific to our solutions, but also occur for the D1-D5-P systems
of~\cite{Liu,brecher}. Curiously, one cannot pass to the covering
space where $u$ is a non-compact coordinate as $z_5$ has to be
periodic when $q^3>0$ in order to avoid a Dirac-Misner string
singularity at $x=1$~\cite{Elvang}. Thus unlike the BMPV case we
cannot remove these CTC, unless $q^3=0$ in which case the horizon is
singular.

The microscopic derivation of the entropy of the asymptotically flat
supertubes has been examined in~\cite{bkraus}, though no complete
derivation has been given in terms of the D1-D5-P CFT. Moreover there has
been success from the eleven dimensional standpoint by considering
the CFT of the M5 brane intersection~\cite{cyrier}. It is
interesting to note that since the geometry of the horizon is
unaffected by the deformation we introduced, the entropy of this
configuration is the same as that of the asymptotically flat three
charge supertubes found in~\cite{3chargestube}. Now, strings in the
Penrose limit of $AdS_3 \times S^3 \times T^4$ can be easily
quantized~\cite{russo}. It turns out that the asymptotic density of
states for this model is the same as in flat space (as $\alpha' \to 0$). This is nice as
it suggests that given the microscopic derivation of the entropy of
the three charge supertube in flat space one should find the same
answer as for the three charge supertube in the plane wave solution
constructed in this paper. This is consistent with the fact that the
Bekenstein-Hawking entropy is unaffected by the plane wave as
remarked above.

\section{Supertubes in the maximally supersymmetric plane wave background}
The solution we have constructed may be reduced along $S^1 \times
T^4$ to yield the seed G\"odel solution we started with. This of
course requires the metric to be independent of the compactification
coordinates. On the other hand, one may wish to consider supertube
configurations embedded in more general plane wave backgrounds, such
as the maximally supersymmetric plane wave background.
\par
Consider the undeformed solution. The Einstein frame metric is given
in terms of the string frame metric as ${(g_E)}_{\mu\nu}
=e^{-\phi/2} {(g_S)}_{\mu\nu}$. For the case at hand \bea \nonumber
ds^2_E &=& H_1^{-1/4} H_2^{-3/4} \Big[ 2(dt+\omega)(dz_5+\Omega) +
H_3(dz_5+\Omega)^2 \Big]
\\&+&H_1^{3/4} H_2^{1/4} \sum_{i=1}^4 dx_i^2 + \left( \frac{H_2}{H_1}
\right)^{1/4} \sum_{i=1}^4 dz_i^2, \\ \Omega &\equiv&
-\frac{q^3}{2}( (1+x)d\phi + (1+y) d\psi). \eea Changing
variables\footnote{Note that in this section our $u,v$ coordinates are
  defined differently to the previous section, in order to match with~\cite{brecher}.}
to $2u=-z_5$ and $2v=2t+z_5$ transforms the terms in the square
bracket to \be -4dudv -2du(2\omega+\Omega) +(2dv+2\omega)\Omega
+\Omega^2 +(H_3-1)(2du-\Omega)^2. \ee This can be cast in a nicer
form by introducing a new $u$ coordinate by $du \to du +\Omega/2$
which leaves the full metric as \bea \nonumber ds^2_E &=& H_1^{-1/4}
H_2^{-3/4} \Big[ -4dudv - 2du(2\omega+\Omega) +4(H_3-1)du^2 \Big]
\\&+&H_1^{3/4} H_2^{1/4} \sum_{i=1}^4 dx_i^2 + \left( \frac{H_2}{H_1}
\right)^{1/4} \sum_{i=1}^4 dz_i^2. \eea This is rather nice as we
have cast the D1-D5-P supertube in exactly the same form as the
D1-D5-P system which reduces to the BMPV black hole. This form of
the metric allows one to add in extra pieces to the $g_{uu}$
component very easily as was done for the BMPV system
in~\cite{brecher}. This relies on the following observation: \bea
ds^2 &=& e^{2A} \left(-4dudv + \mathcal{H} du^2 + du \sum_{i=1}^n
C_i dx_i \right) + \sum_{i=1}^n e^{2B_i}dx_i^2
\\ \Rightarrow R_{\mu\nu} &=& \bar{R}_{\mu\nu} - \frac{1}{2}
\delta^u_{\mu} \delta^{u}_{\nu} e^{2A} \sum_{i=1}^n e^{-2B_i} [
\partial_i^2 \mathcal{H} + \partial_i \mathcal{H} \partial_i G_i+ 2(\partial_i^2A +\partial_i A \partial_i G_i)\mathcal{H}] \\ G_i &=& 2A- 2B_i
+\sum_{j=1}^n B_j \eea where $A,B_i,C_i$ are all functions of the
transverse coordinates $x_i$ and $\bar{R}_{\mu\nu}$ is the Ricci
tensor with $\mathcal{H}=0$\footnote{In~\cite{brecher} it was stated
that the inclusion of angular momentum does not affect the result
quoted in~\cite{Liu} which had $C_i=0$.}.

Armed with these results we now deform the D1-D5-P supertube as
follows: \bea \nonumber ds^2_E &=& H_1^{-1/4} H_2^{-3/4} \Big[
-4dudv - 2du(2\omega+\Omega) +(\mathcal{H}+4(H_3-1))du^2 \Big]
\\&+&H_1^{3/4} H_2^{1/4} \sum_{i=1}^4 dx_i^2 + \left( \frac{H_2}{H_1}
\right)^{1/4} \sum_{i=1}^4 dz_i^2. \eea In view of the general
result quoted above it is immediate that the the Ricci tensor of
this deformed configuration is: \bea R_{\mu\nu} = \bar{R}_{\mu\nu} -
\frac{1}{2H_1H_2} \delta^u_{\mu} \delta^u_{\nu} \left(
\partial_{x_i}^2 + H_1 \partial_{z_i}^2 - \frac{1}{4}\left( \partial_{x_i}^2\log H_1 + 3\partial_{x_i}^2\log H_2 \right) \right) \mathcal{H}. \eea It should be noted that in this case the functions $G_i=0$ and thus there are no terms with first derivatives of $\mathcal{H}$. We will choose to support this
deformation by a five form flux as in~\cite{brecher}. Note that this
is in contrast to the previous section where the deformation was
supported by a three form flux. The five form flux must be self dual
and $F_5 \wedge F_3=0$. It is natural to try and use the same
expression as in~\cite{brecher}. Namely \be F_5 = \mu du
\wedge ( dx_1 \wedge dx_2 \wedge dz_1 \wedge dz_2 +dx_3 \wedge dx_4
\wedge dz_3 \wedge dz_4). \ee It is in fact straightforward to check
that this five form is self dual also in this case. However in
general $F_3 \wedge F_5 \neq 0$, see Appendix. In the special case $q^3=0$ though
$F_3 \wedge F_5=0$ and thus we may straightforwardly deform the
solution in this situation. Note that the supertube is nakedly
singular in this limit though. Nevertheless as noted
in~\cite{3chargestube} the world volume theory of this
configuration has a sensible interpretation in Type IIA in terms of
D6 branes.
\par The analysis follows similarly to~\cite{brecher}. Explicitly, the Type IIB Einstein
equations are
\begin{equation}
R_{\mu\nu} = \frac{1}{2}\partial_{\mu} \phi \partial_{\nu} \phi +
\frac{1}{96}F_{\mu \alpha \beta \gamma
  \delta}F_{\nu}^{\phantom{a}\alpha \beta \gamma \delta} +
\frac{e^{\phi}}{4}\left (F_{\mu \alpha
  \beta}F_{\nu}^{\phantom{b}\alpha \beta} -
\frac{1}{12}g_{\mu \nu}F^{2}_{3} \right )
\end{equation}  Clearly only the $uu$ component of the stress energy
tensor is altered by the presence of $\mathcal{H}$. It is not
difficult to check that these extra terms are
\bea
\Delta (F_{\mu \alpha \beta}F_{\nu}^{\phantom{b} \alpha \beta}) &=& -2(\partial_{i}\log{H_{2}})^2
H_{1}^{-\frac{1}{2}}H_{2}^{-\frac{3}{2}}\mathcal{H} \; \delta^u_{\mu} \delta^u_{\nu} \\
 \Delta(g_{\mu \nu}F^2_{3}) &=& 6((\partial_{i}\log{H_{1}})^2 -
 (\partial_{i}\log{H_{2}})^2) H_{1}^{-\frac{1}{2}}H_{2}^{-\frac{3}{2}}
 \mathcal{H} \; \delta^u_{\mu} \delta^u_{\nu}\\
F_{\mu \alpha \beta \gamma \delta}F_{\nu}^{\phantom{a}\alpha \beta
  \gamma \delta} &=& \frac{48 \mu^2}{H_1 H_2}
\delta^u_\mu \delta^u_\nu \eea where $\Delta$ represents the change in
the quantity in the brackets due to the deformation $\mathcal{H}$. The
equations of motion for the dilaton and $F_3$ are unchanged.
Enforcing the $uu$ component of the Einstein equations then leads to
a simple linear equation for the deformation $\mathcal{H}$: \be
\label{PDE} (\partial_{x_i}^2 + H_1 \partial_{z_i}^2) \mathcal{H} =
-\mu^2. \ee Note that to do this one needs to use the fact that
$H_1,H_2$ are harmonic when $q^3 = 0$.  Thus the equation for the
deformation is identical to that found in~\cite{brecher} for the
BMPV system, except now $H_1$ is a harmonic function with delta
function sources on a ring $\rho=R$ and $\theta=\pi/2$ as opposed to
at the origin. This makes the PDE (\ref{PDE}) more complicated and
in fact if one writes it in $(\rho,\theta)$ or $(x,y)$ coordinates
it is not separable. Remarkably, if one uses yet a different
coordinate system the equation can be made separable. The
coordinates in question are: \be r^2 = R^2\frac{1-x}{x-y}, \qquad
\cos^2\Theta = \frac{1+x}{x-y} \ee and were introduced
in~\cite{3chargestube}. The flat space metric then looks like \be
\sum_{i=1}^4 dx^2_i = \Sigma \left( \frac{dr^2}{r^2+R^2} + d\Theta^2
\right) + (r^2+R^2) \sin^2\Theta d\psi^2 + r^2\cos^2\Theta d\phi^2
\ee where $\Sigma = r^2 +R^2\cos^2\Theta$. It is immediately
apparent that there are solutions of the form $\mathcal{H} = X(x_i) + Z(z_i)$ where
$\partial_{z_i}^2 Z =\alpha^2$ where $\alpha^2$ is a separation
constant. The resulting equation for $X$ is then $\partial_{x_i}^2X
+\alpha^2 H_1 = -\mu^2$, which upon multiplication by $\Sigma$ is
additively separable in the $r,\Theta$ coordinates system since
$H_1=1+Q_1/\Sigma$. This means that $X= F(r)+ G(\Theta)$. The function
$F(r)$ satisfies \bea
\frac{1}{r} \frac{d}{dr} \left( r(r^2+R^2)\frac{dF}{dr}\right) +r^2(\alpha^2+\mu^2) &=&
-\beta^2 
\eea where $\beta^2$ is another separation constant. This may be be integrated to give \bea \nonumber F(r)
&=& -\frac{r^2}{8}(\alpha^2+\mu^2) + \frac{R^2}{8}(\alpha^2 +\mu^2)
\log(r^2+R^2) -\frac{\beta^2}{4} \log(r^2+R^2) \\&+& c_1 \log(
r/\sqrt{r^2+R^2} ) +c_2 \eea where $c_1,c_2$ are integration
constants. The equation for $G$ may also be integrated, but it is
convenient to change variables to $z=\sin^2\Theta$ first. In terms
of $z$ we have \be 4 \frac{d}{dz}\left( z(1-z) \frac{dG}{dz} \right) + R^2(\alpha^2
+\mu^2)(1-z) +Q_1\alpha^2 =\beta^2 \ee which leads to \bea G(z) &=&
\frac{(Q_1 \alpha^2 - \beta^2)}{4}\log(1-z) + \frac{R^2(\alpha^2 +
\mu^2)}{8}(\log(1-z) - z) \\ &+& c_3(\log{z} - \log(1-z)) + c_4.
\eea We have generated quite a few constants upon integrating
(\ref{PDE}), however they may all be fixed as follows. The constants
$c_2,c_4$ can be absorbed into shifts of $v$. Demanding
regularity\footnote{The horizon in these coordinates is located at
$r=0$ and $\Theta=\pi/2$, however as already noted it is not regular
in the undeformed solution since $q^3=0$.} at $\Theta=0$ and $\pi/2$
forces $c_3=0$ and $\beta^2=Q_1\alpha^2+ R^2(\alpha^2+\mu^2)/2$
respectively. Demanding regularity at $r=0$ forces $c_1=0$ and
finally requiring that $\mathcal{H} \sim \const (x_ix_i +z_iz_i)$
tells us that $\alpha^2=-\mu^2/2$. Thus we arrive at \be \mathcal{H}
= -\frac{\mu^2}{16} (r^2 +R^2\sin^2\Theta +z_iz_i)
+\frac{1}{8}Q_1\mu^2 \log(r^2+R^2). \ee This deformation gives a
three charge, two dipole supertube which asymptotes to the maximally
supersymmetric plane wave solution of type IIB supergravity. The
remarks concerning the global causal structure of these deformed
supertube configurations given in the previous section also apply
here, however in this case we do not get the quantization condition on
$\mu$. In particular, we can pass to the covering space where $u$ is
non-compact. One would expect in this case the solution to be devoid
of CTC (given the constraint on the charges in~\cite{3chargestube}), though we have not performed a full analysis in the
intermediate region (in between the near horizon and the asymptotic
plane wave). 

Finally we note
in passing there is a class of solutions (with $\alpha = 0$) that are independent of the
toroidal directions $z_i$; in this case, one could easily compactify
on $S^1 \times T^4$ to derive nakedly singular asymptotically
G\"odel spacetimes in five dimensional minimal supegravity coupled
to two vector multiplets.

\section{Concluding remarks}
We have demonstrated  how three charge supertubes can be embedded
not only in asymptotically flat backgrounds, but also in the next
simplest class of solutions, supersymmetric plane wave spacetimes.
This was shown explicitly in two cases: firstly, for  the Penrose
limit of $AdS_3 \times S^3 \times T^4$, and secondly for the
maximally supersymmetric BFHP plane wave. The former case was
derived by first deforming the supersymmetric black rings in D=5
minimal supergravity coupled to $N-1$ vector multiplets. In the case
$N=3$, this solution was uplifted to eleven dimensions to describe
three charge supertubes in the supersymmetric G\"odel background.
Upon Kaluza Klein reduction and T dualisation, a D1-D5-P supertube
embedded in $\frac{1}{2}$ BPS plane wave was constructed.
Furthermore, the inclusion the plane wave term does not affect the
properties of the horizon. The removal of Dirac-Misner string
singularities, however, leads to CTC, in contrast to the BMPV
case.  
\par In the second case, we use a more direct approach to derive a supertube plane wave configuration supported by a self dual five form
flux. In order to satisfy the equations of motion, it seems one has
to turn off the Kaluza-Klein dipole charge. By noting that the Ricci
tensor has a simple decomposition under wave-like deformations of
the metric we arrive at a simple linear PDE for the
deformation. We show that this is additively separable, in suitable
coordinates, and under this condition derive the general solution. A particular solution
corresponds to supertubes in the maximally supersymmetric
background.
\par
There remain several open problems concerning asymptotically plane
wave supertubes. Obviously, one might be interested in their world
volume description. This could be particularly relevant for the
solutions describing supertubes in the maximally supersymmetric
background, as they seem to be nakedly singular from their
supergravity description. Furthermore, one could try to generalise
the solution presented here in the maximally supersymmetric plane wave background such that one
has a non-zero Kaluza Klein dipole charge. Since these three charge,
two dipole supertubes have non-extremal
counterparts~\cite{nonextrem} with regular horizons, it might be
useful to consider plane wave extensions of these thermally excited
supertubes first.
\par
Finally we should emphasise that it is most interesting that one can
embed such supertube configurations in non-trivial backgrounds such
as plane waves and G\"odel spacetimes. These arose from lifting
G\"odel black rings. There remains, however, the interesting and
apparently difficult problem of finding asymptotically AdS black rings.

\acknowledgments We would like to thank Roberto Emparan for useful
comments and reading through a draft of the paper. HKK would like to
thank St. John's College, Cambridge, for financial support.

\appendix
\section{The three form $F_3$}
The two form field strengths of the $U(1)^N$ five dimensional
supergravity are given by: \be F^I = dA^I = d(X^Ie^0)+ \Theta^I =
f^{-1} dH_I^{-1} \wedge e^0 + H_I^{-1} d\omega + \Theta^I . \ee It
then follows that the hodge dual is \be *_5 F^I = \frac{1}{3!}
f^{-2} \partial_i H_I^{-1} \epsilon_{ilmn} dx^l\wedge dx^m \wedge
dx^n + H_1^{-1} f^{-1} e^0 \wedge (G^+-G^-) + e^0 \wedge \Theta^I
\ee where we have used the fact that $*_4 d\omega = f^{-1}(G^+-G^-)$
and $*_4 \Theta^I =\Theta^I$. Note that $\epsilon_{0x_1x_2x_3x_4}=1$
has been used in these Cartesian coordinates. The three form in IIB
is given by \be F_3 = (X^1)^{-2} *_5 F^1 + F^2\wedge (dz_5 +A^3) =
(X^1)^{-2} *_5 F^1- 2F^2 \wedge du + F^2\wedge (A^3 -\Omega),
\ee
where we have used the new $u$ coordinate defined in section 4. Thus we will also need \bea F^2 \wedge (A^3 -\Omega) = f^{-1}
H_2^{-1}H_3^{-1} d\omega \wedge e^0
+ f^{-1} H_3^{-1} \Theta^2 \wedge e^0 \\
 = \frac{H_1^{1/3}}{H_2^{2/3} H_3^{2/3}} d\omega \wedge e^0 +
 \frac{H_1^{1/3} H_2^{1/3}}{H_3^{2/3}} \Theta^2 \wedge e^0 \eea
and \bea \nonumber (X^1)^{-2} *_5 F^1 = \frac{1}{3!} f^{-2}
(X^1)^{-2}\partial_i H_1^{-1} \epsilon_{lmni} dx^l\wedge dx^m \wedge
dx^n \\+ \frac{H_1^{1/3}}{H_2^{2/3} H_3^{2/3}} e^0 \wedge
f^{-1}(G^+-G^-) + (X^1)^{-2}e^0 \wedge \Theta^1. \eea Upon adding
A.5 to A.6 we see that the terms that look like $e^0 \wedge G^-$
cancel. Interestingly we find further cancellations. Using the
identity $G^+= -\frac{1}{2}f H_I \Theta^I$ leads to \bea
\frac{H_1^{1/3}}{H_2^{2/3}H_3^{2/3}} 2 e^0 \wedge f^{-1}G^+ +
(X^1)^{-2}e^0 \wedge \Theta^1 + f^{-1}H_3^{-1} e^0 \wedge \Theta^2 =
-f^{-1}H_2^{-1} e^0 \wedge \Theta^3. \eea Putting everything
together the final explicit expression for $F_3$ is: \bea F_3
=-2F^2\wedge du +\frac{1}{3!} f^{-2} (X^1)^{-2}\partial_i
H_1^{-1} \epsilon_{ilmn} dx^l\wedge dx^m \wedge dx^n -f^{-1}H_2^{-2}
e^0 \wedge \Theta^3. \eea Thus as compared to the D1-D5-P BMPV
system of~\cite{brecher} we have an extra term $-f^{-1}H_2^{-2} e^0
\wedge \Theta^3$. Note that this term is proportional to $q^3$ the
KK dipole charge. Thus, shutting off the KK dipole charge of the
supertube gets rid of this extra term leaving $F_3 \wedge F_5=0$ as
required.

\end{document}